\documentclass[]{emulateapj}

\begin{document}

\title{Can Self--Organizing Maps accurately predict photometric redshifts?}

\author{M.J. Way\altaffilmark{1,2,3},
C.D. Klose\altaffilmark{4}}

\altaffiltext{1}{NASA Goddard Institute for Space Studies, 2880 Broadway,
New York, New York 10025, USA}
\altaffiltext{2}{NASA Ames Research Center, Space Sciences Division, MS 245--6,
Moffett Field, California 94035, USA}
\altaffiltext{3}{Department of Astronomy and Space Physics, Uppsala, Sweden}
\altaffiltext{4}{Think Geohazards, 205 Vernon Street, Suite A Roseville, CA 95678, USA}

\begin{abstract}
We present an unsupervised machine learning approach that can be employed for
estimating photometric redshifts. The proposed method is based on a vector
quantization approach called Self--Organizing Mapping (SOM).  A variety of
photometrically derived input values were utilized from the Sloan Digital
Sky Survey's Main Galaxy Sample, Luminous Red Galaxy, and Quasar samples
along with the PHAT0 data set from the PHoto-z Accuracy Testing project.
Regression results obtained with this new approach were evaluated in terms
of root mean square error (RMSE) to estimate the accuracy of the
photometric redshift estimates.  The results demonstrate competitive RMSE
and outlier percentages when compared with several 
other popular approaches such as Artificial Neural Networks and Gaussian
Process Regression.  SOM RMSE--results (using $\Delta$z=z$_{phot}$--z$_{spec}$)
for the Main Galaxy Sample are 0.023,
for the Luminous Red Galaxy sample 0.027, Quasars are 0.418, and PHAT0 synthetic
data are 0.022. The results demonstrate that there are non--unique solutions
for estimating SOM RMSEs.  Further research is needed in order to find more
robust estimation techniques using SOMs, but the results herein are a
positive indication of their capabilities when compared with other well-known
methods.
\end{abstract}


\section{Introduction}

There is a pressing need for accurate estimates of galaxy photometric
redshifts (photo--z's) as demonstrated by the increasing number of papers
on this topic and especially by recent attempts to objectively compare
methods \citep[e.g.][]{Hildebrandt2010,Abdalla2011}.
The need for photo-z's will only increase as larger and deeper
surveys such as Pan-STARRS\footnote{Panoramic Survey Telescope \& Rapid Response System}\citep{Kaiser2004},
LSST\footnote{Large Synoptic Survey Telescope}\citep{Ivezic08}
and Euclid \citep{Sorba2011} come on--line in the coming decade. The
photometric--only surveys (Pan-STARRS, LSST) will have relatively small numbers of
follow-up spectroscopic redshifts and will rely upon either
template-fitting methods such as Bayesian Photo-z's \citep{Benitez2000}
Le Phare \citep{Ilbert2006}, or training-set methods such as those
discussed herein. The Euclid mission may include a slitless spectrograph
offering far more training--set galaxies.

A diverse set of regression techniques using training--set methods have been
applied to the problem of estimating photometric redshifts in the past 10 years.
These include Artificial Neural Networks
\citep{Firth03,Tagliaferri03,Ball04,CL2004,Vanzella04},
Decision Trees \citep{Suchkov05}, Gaussian Process Regression
\citep{Way06,Foster09,Way09,Bonfield2010,Way2011}, Support Vector Machines
\citep{Wadadekar05}, Ensemble Modeling \citep{Way09}, Random Forests
\cite{Carliles08}, and Kd--Trees \citep{Csabai03} to name but a few.

On the other hand, even though Self--Organizing Maps (SOMs) have been
used extensively in a number of other scientific fields (the paper that
opened the field, \cite{Kohonen1982}, currently has over 2000 citations)
they have been used sparingly thus far in
Astronomy \citep[e.g.][]{Mahdi2011,Naim1997,WGS2011}, and only this year
in estimating photometric redshifts \citep{geach2011}.

In this work we attempt to use SOMs to estimate photometric
redshifts for several Sloan Digital Sky Survey \citep[SDSS,][]{York2000}
derived catalogs of different galaxy types, including Quasars along with the PHAT0 data set
of \cite{Hildebrandt2010}.  In Section \ref{sec:data} we describe
the input data sets used, in Section \ref{sec:methods} we give an overview
of SOMs, and some conclusions in Section \ref{sec:conclusion}.

\section{Data}\label{sec:data}

Three different data sets derived from the SDSS
Data Release Seven \citep[DR7,][]{DR7} were used.
They include the Main Galaxy Sample \citep[MGS,][]{Strauss02}
the Luminous Red Galaxy Sample \citep[LRG,][]{Eisenstein01},
and the Quasar sample \citep[QSO,][]{Schneider2007}.
Data from the Galaxy Zoo\footnote{http://www.galaxyzoo.org} \citep{Lintott2008}
Data Release 1 \citep{Lintott2011} survey results were used
to segregate galaxies as Spiral or Elliptical in the case of the MGS and
LRG samples. Details of how this was done are given in \cite{Way2011}.
Dereddened magnitudes (u,g,r,i,z) were used as inputs in all scenarios.
The same SDSS photometric and redshift quality flags on the input variables
were used as in \cite{Way2011}. 
In addition we used the simulation--based PHAT0 data set
\citep[see][]{Hildebrandt2010} which was
constructed to to test a variety of different photo--z estimation methods.
The PHAT0 data set consists of 5 SDSS like filters (u,g,r,i,z) used on
MEGACAM at CFHT \citep{Boulade2003} with an additional 6 input filters
(Y,J,H,K,Spitzer IRAC [3.6], Spitzer IRAC [4.5])
giving a total of 11 filters spanning a range of 4000{\AA} to 50,000{\AA}.
This large range should help to avoid color--redshift degeneracies
that can occur if ultraviolet or infrared bandpasses are not
used \citep{Benitez2000}.  The PHAT0 synthetic photometry was
created from the Le Phare photo-z code \citep{Arnouts2002,Ilbert2006}.
Initially Le Phare creates noise free data, but given the desire
to test more real--world conditions we utilized the PHAT0 data
with added noise. A parametric form was used for the signal--to--noise
as a function of magnitude where it acts as an exponential at
fainter magnitudes and a power--law a brighter ones. The magnitude cut
between these two regimes is filter dependent and is given in Table 2
of \cite{Hildebrandt2010}. The larger of two catalogs was used herein
(as suggested for training--set methods) that contains $\sim$ 170,000 objects.

Since we use a training--set method our original data sets are
split into training=89\%, testing=10\% and validation=1\%.
Validation was only used in the Artificial Neural Network algorithm
discussed in the next section. The full size
of each input data set are listed in parentheses in column 1
of Table \ref{tbl-1}.

\section{Methods}\label{sec:methods}

Several methods in use for calculating photometric
redshifts were compared with the SOM results: the Artificial Neural Network
code of \cite{CL2004} (ANNz), the Gaussian Process Regression code
of \cite{Foster09} (GPR), as well as simple Linear and Quadratic regression.
The latter is comparable to that of the Polynomial fits used by \cite{LiYee2008}.
Both the ANNz and GPR codes are freely
downloadable\footnote{GPR: http://dashlink.arc.nasa.gov/algorithm/stableGP and
ANNz: http://www.star.ucl.ac.uk/~lahav/annz.html}. Details on the ANNz and GPR
algorithms can be found in their respective citations above. 

\begin{deluxetable}{llcccc}
\tabletypesize{\tiny}
\tablecolumns{6}
\tablecaption{Results\label{tbl-1}}
\tablehead{
\colhead{Data\tablenotemark{a}}   & \colhead{Method\tablenotemark{b}} &                & \colhead{$\sigma_{RMSE}$\tablenotemark{c}}&                &\colhead{Outlier\tablenotemark{d}}\\
                                  &                                   & \colhead{50\%} & \colhead{10\%}                            & \colhead{90\%} &                                  }
\startdata
MGS     &GPR    & 0.02087 & 0.02072 & 0.02096&0.11629\\
(455803)&ANNz   & 0.02044 & --       & --      &0.14482\\
--       &SOM    & 0.02339 & --       & --      &0.1689\\
--       &Linear   & 0.02742 & 0.02729 & 0.02758  & 0.35986\\
--       &Quadratic& 0.02494 & 0.02412 & 0.02762  & 0.29184\\
\tableline
LRG      &GPR      & 0.02278 & 0.02256 & 0.02309  & 0.41898\\
(143221) &ANNz     & 0.02138 & --      & --       & 0.41176\\
--       &SOM      & 0.02689 & --      & --       & 0.64292\\
--       &Linear   & 0.02896 & 0.02896 & 0.02897  & 0.71516\\
--       &Quadratic& 0.02382 & 0.02376 & 0.02402  & 0.45510\\
\tableline
MGS--ELL &GPR      & 0.01455 & 0.01434 & 0.01473  & 0.06591\\
(45521)  &ANNz     & 0.01442 & --      & --       & 0.06591\\
--       &SOM      & 0.02044 & --      & --       & 0.10984\\
--       &Linear   & 0.01745 & 0.01731 & 0.01756  & 0.19772\\
--       &Quadratic& 0.01612 & 0.01609 & 0.01629  & 0.10984\\
\tableline
MGS--SP  &GPR      & 0.02078 & 0.02061 & 0.02093 & 0.13305\\
(120266) &ANNz     & 0.01991 & --      & --      & 0.05821\\
--       &SOM      & 0.02426 & --      & --      & 0.04158\\
--       &Linear   & 0.02539 & 0.02529 & 0.02555 & 0.28272\\
--       &Quadratic& 0.02326 & 0.02296 & 0.02607 & 0.20788\\
\tableline
LRG--SP  &GPR      & 0.01416 & 0.01397 & 0.01436 & 0.00000\\
(13708)  &ANNz     & 0.01516 & --      & --      & 0.00000\\
--       &SOM      & 0.01848 & --      & --      & 0.07299\\
--       &Linear   & 0.01635 & 0.01627 & 0.01649 & 0.07299\\
--       &Quadratic& 0.01469 & 0.01462 & 0.01477 & 0.00000\\
\tableline
LRG--ELL &GPR      & 0.01186 & 0.01162 & 0.01224 & 0.00000\\
(27378)  &ANNz     & 0.01298 & --      & --      & 0.10961\\
--       &SOM      & 0.01568 & --      & --      & 0.00000\\
--       &Linear   & 0.01362 & 0.01361 & 0.01364 & 0.10961\\
--       &Quadratic& 0.01263 & 0.01254 & 0.01274 & 0.07307\\
\tableline
QSO      &GPR      & 0.37342 & 0.03967 & 0.37626 &50.96627\\
(56923)  &ANNz     & 0.65802 & --      & --      &88.54533\\
--       &SOM      & 0.41821 & --      & --      &54.23401\\
--       &Linear   & 0.57061 & 0.57010 & 0.57102 &84.64512\\
--       &Quadratic& 0.53972 & 0.53679 & 0.54539 &81.27196\\
\tableline
phat0    &GPR      & 0.01487 & 0.01436 & 0.01532 & 0.03539\\
(169520) &ANN      & 0.01805 & --      & --      & 0.05309\\
--       &SOM      & 0.02236 & --      & --      & 0.37754\\
--       &Linear   & 0.08703 & 0.08702 & 0.08704 &19.34875\\
--       &Quadratic& 0.02436 & 0.02433 & 0.02438 & 0.19467\\
\enddata
\tablenotetext{a}{MGS=Main Galaxy Sample \citep{Strauss02},
LRG=Luminous Red Galaxies \citep{Eisenstein01}, SP=Classified as spiral
by Galaxy Zoo, ELL=Classified as elliptical by Galaxy Zoo,
QSO=Quasar sample \citep{Schneider2007}}
\tablenotetext{b}{GPR=Gaussian Process Regression \citep{Foster09},
ANNz=Artificial Neural Network \citep{CL2004}, SOM=Self--Organizing Maps 
(SOM-4100 and SOM-5100 see Figure \ref{fig2} for details),
phat0=PHAT synthetic sample}
\tablenotetext{c}{We quote the bootstrapped 50\%, 10\% and 90\% confidence
levels as in \cite{Way09} for the root mean square error (RMSE) when available.}
\tablenotetext{d}{Percentage of points defined as outliers. Following a
prescription similar to that of \cite{Hildebrandt2010} we quote the percentage of points
outside of $\Delta$z=z$_{phot}$--z$_{spec}$ $\pm$ 0.1}
\end{deluxetable}

The main purpose of Self--Organized mapping is the ability of SOMs to transform 
a feature vector of arbitrary dimension drawn from the given feature space 
of photometric inputs (e.g., the SDSS u,g,r,i,z magnitudes) into simplified 
1-- or 2--dimensional discrete maps. The method was originally developed by
\cite{Kohonen1982,Kohonen2001} to organize information in a logical
manner. This type of machine learning utilizes an unsupervised
learning scheme of vector quantization, known as competitive learning in the
field of neural information processing. It is useful for analyzing complex data
with a--priori unknown relationships that are visualized by the
self-organization process \citep{Kohonen2001}.

\begin{figure}
\includegraphics[scale=0.45]{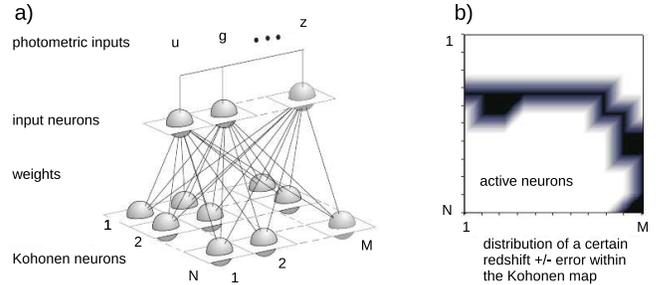}
\caption{Schematic illustration of the structure (a) and functionality (b)
of a Self--Organizing Map with $I$ input neurons and $M \times N$ Kohonen
neurons. The SOM visualizes the structure of the $I$--dimensional input space.
In this case, the SOM illuminates a certain redshift$\pm$error within the
Kohonen map and as a function of the input space.}\label{fig1}
\end{figure}

A SOM is structured in two layers: an input layer and a Kohonen layer
(Figure \ref{fig1}). For example, the Kohonen layer could represent a
structure with a single 2--dimensional map (lattice) consisting of
neurons arranged in rows and columns.  Each neuron of this discrete lattice
is fixed and is fully connected with all source neurons in the input layer.
For the given task of estimating photometric redshifts, a 5--dimensional
feature vector of the u,g,r,i,z magnitudes is defined.
One feature vector (u,g,r,i,z) is presented to 5 input layer neurons.
This typically activates (stimulates) one neuron
in the Kohonen layer.
Learning occurs during the self--organizing procedure as feature 
vectors drawn from a training data set are presented to the input layer of
the SOM network (Figure \ref{fig1}a). 
These feature vectors are also referred to as input vectors. Neurons of the
Kohonen layer compete to see which neuron will be activated by the weight
vectors that connect the input neurons and Kohonen neurons. 
In other words, the weight vectors identify which input vector can represented
by a single Kohonen neuron.  Hence, they are used to determine only
one activated neuron in the Kohonen layer after the winner--takes--all
principle (Figure \ref{fig1}b). 

\begin{figure}
\includegraphics[scale=0.7]{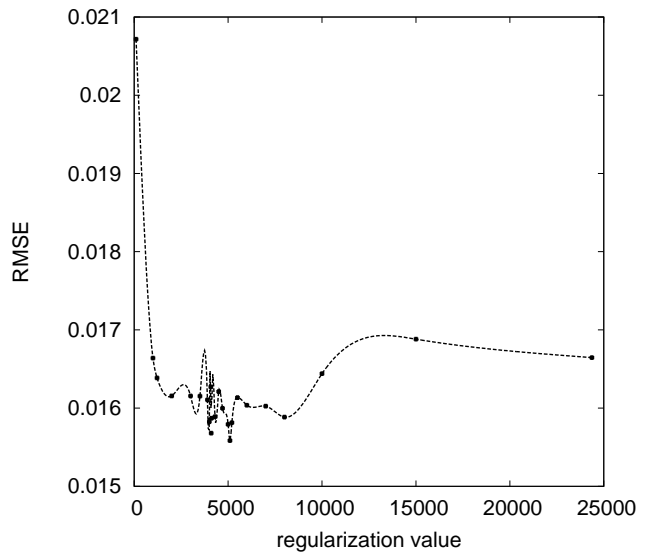}
\caption{Accuracy (RMSE) versus regularization parameter value $\xi$ for the
LRG--ELL data set (see Table \ref{tbl-1}). Different classifications will
result from different choices of the $\xi$ value.
The regularization value is defined by the number of
Kohonen neurons, which is maximum with respect to the training data set. The
convex curve has a two local minima at $\xi$=4100 and $\xi$=5100.
The roughness of this RMSE cost function shows that traditional gradient
based optimization strategies, e.g. deterministic annealing, might result in
sub--optimal solutions. Other methods, such as, genetic programming might find
the global minimum much faster.}\label{fig2}
\end{figure}

The SOM is considered as trained after learning, at which time the weights of the
neurons have stored the inter--relations of all 5--dimensional u,g,r,i,z feature
vectors. Then, known spectroscopic redshift values for all input vectors
of a test data set that are represented by a single Kohonen neuron are averaged
(Fig.\ref{fig1}b).  The redshift mean value represents all 5--D u,g,r,i,z
vectors that are similar to the weight vector of the activated Kohonen neuron. 
The more Kohonen neurons there are the more precisely each input vector can
be represented by a weight vector. However, the total number of Kohenen neurons
are optimized for each data set (see Figure \ref{fig2}).  A practical overview about the
learning/training process is described
by \cite{Klose2006,Klose-etal2008,Klose-etal2010} and in much greater
detail by \cite{Kohonen2001}.

After training, the u,g,r,i,z input vectors of a test data set are presented
to a trained SOM. At the end of a classification step, every Kohonen neuron
approximates an input vector whereby similar/dissimilar input data were
represented by neighboring/distant neurons. One neuron could even classify
several input vectors,
if these input vectors were very similar compared to
the other given input vectors. Results from the photometric redshift
approximations are then compared to known spectroscopic redshift data. 
Regression performance is estimated based on the root mean square error
(RMSE) of the predicted photometric redshifts and the known spectroscopic
redshifts (using $\Delta$z=z$_{phot}$--z$_{spec}$). To reiterate,
during the training phase, each Kohonen neuron
identifies a certain number of galaxies that are characterized by
similar u,g,r,i,z intensities.  Photometric redshift data were then
averaged for these intensity values.

The SOM approximates the input feature
space and maps it into an output space. The output space shows the
SOM approximation as a 2-D map \citep{Haykin2009}.
Best results can be obtained with an optimization scheme such that the
RMSE of the test data set is minimal as illustrated in Figure \ref{fig2}. 
Accuracy (e.g. RMSE) depends on the size of the Kohonen map.
The number of neurons in the Kohonen map can be considered
a regularization parameter ($\xi$) as shown in Figure \ref{fig2}.

Figure \ref{fig2} shows that RMSE is high when the number of Kohonen neurons 
is too small ($\xi<$2000) or too large ($\xi>$10000) and hence that
the 5--dimensional u,g,r,i,z--input space is underfit or overfit.  
Theoretically, a global minimum of the RMSE--curve might exist. However, 
the input feature space for the given photometric redshift problem shows a very 
rough RMSE--curve (Figure \ref{fig2}) with at least 2 local minima. In this case
it is clear that SDSS redshift estimation tends to have several local minima,
which makes is important to chose the right optimization method to determine the
SOM network size. On the other hand, the smoother the RMSE--curve is the better
gradient methods can be utilized.  Evolution strategies or genetic programming
could be applied to rougher curves with many local minima. This in turn can
make it cumbersome to find fast back--propagation Artificial Neural Network
(ANN) structures, especially when data sets are small.  

Another advantage of SOMs in comparison to ANNs is that there is no need to
optimize the structure of SOMs (e.g., number of hidden layers), since it is
based on unsupervised learning. 

Only the size of the Kohonen map needs to
be optimized for each data set.  SOMs also allow non--experts to visualize
the redshift estimates in relation to the multi--dimensional input space.
This eliminates the often criticized ``black box" problem of ANNs.
As mentioned previously, SOMs approximate the input feature space while
ANNs typically separate them into sub--regions.
Finally, SOMs are known to be powerful when very small data sets are available for
training \citep[see, ][]{Haykin2009}.

\begin{figure}
\centering
\begin{tabular}{cc}
\includegraphics[scale=0.25]{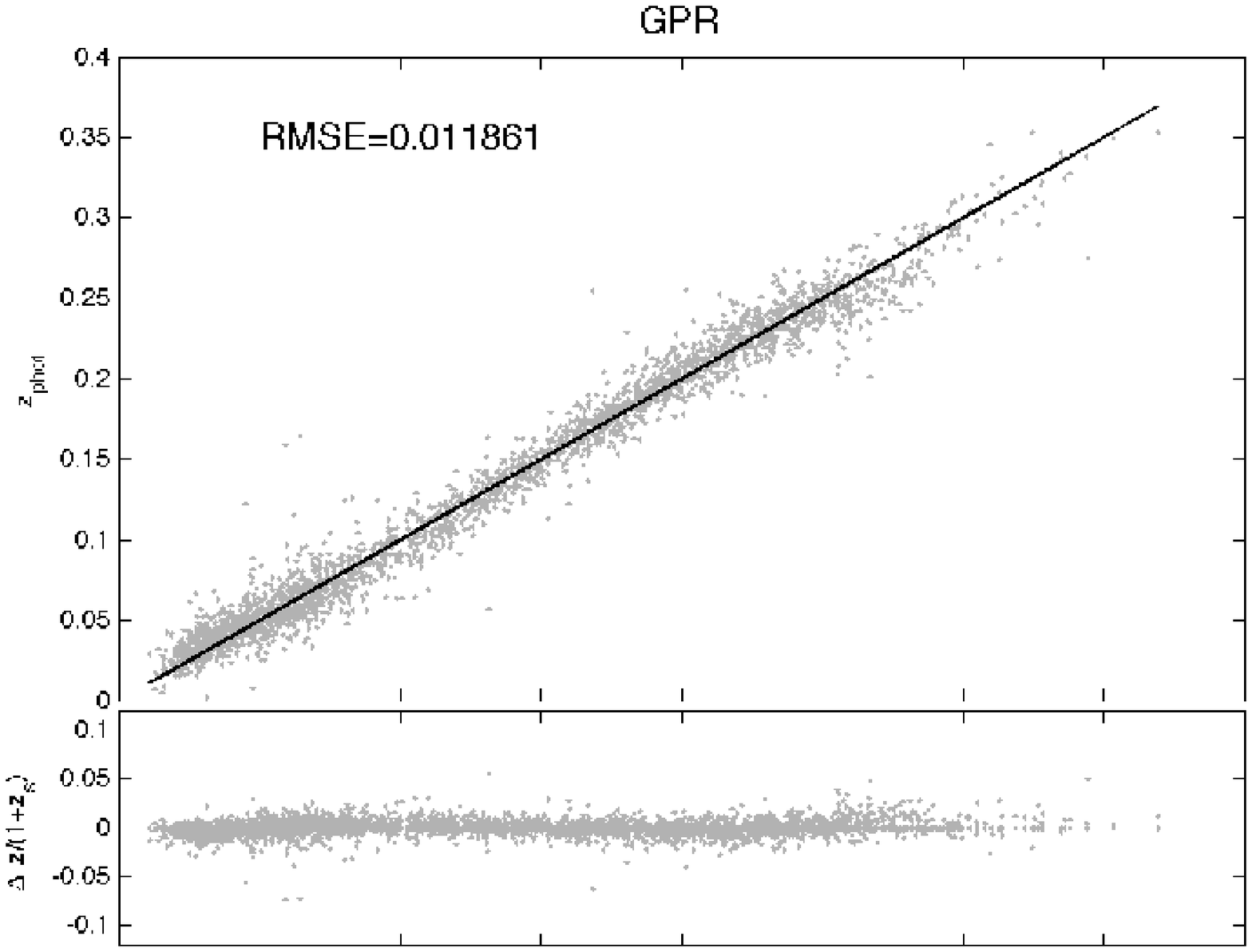} &
\includegraphics[scale=0.25]{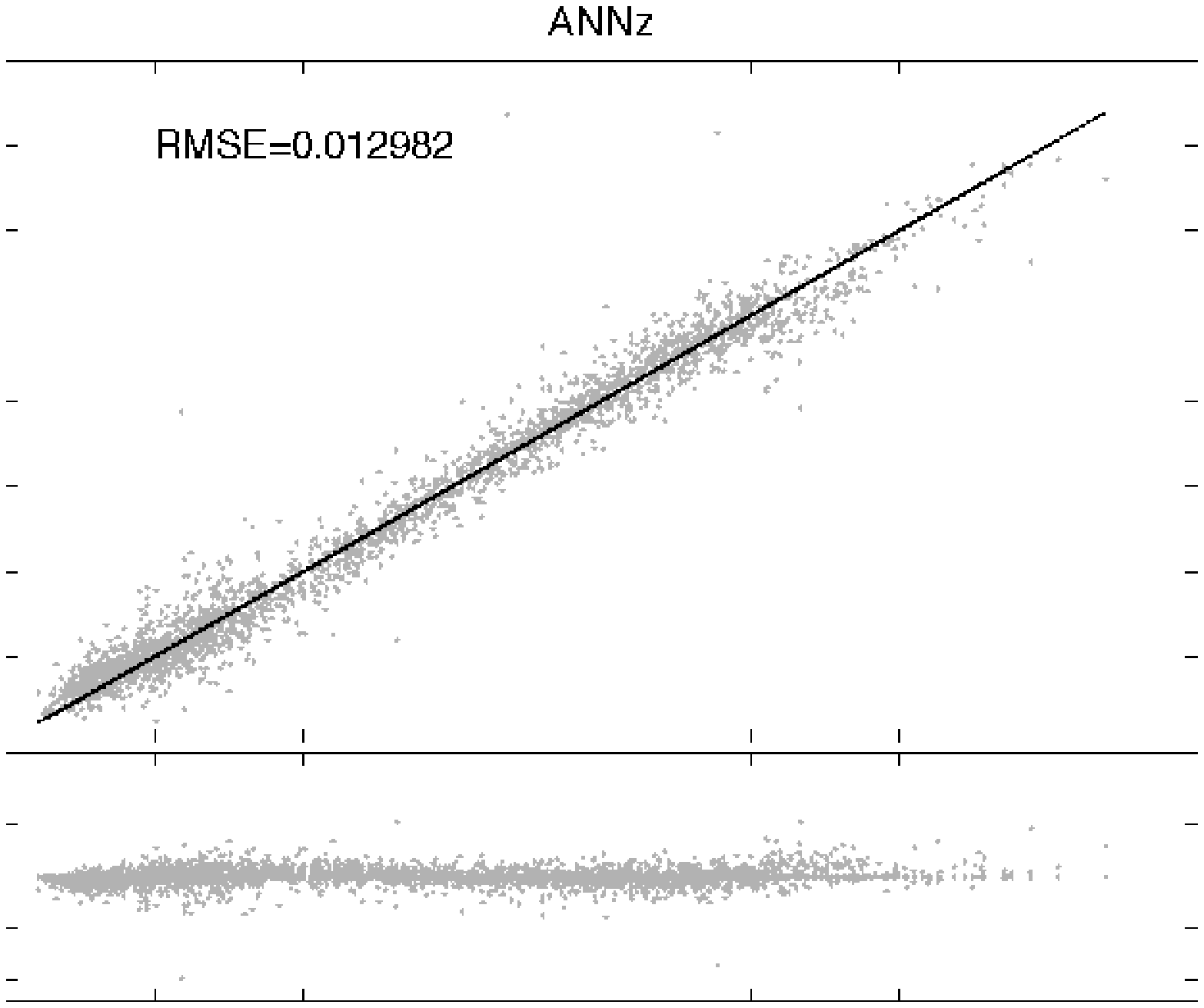} \\
\includegraphics[scale=0.25]{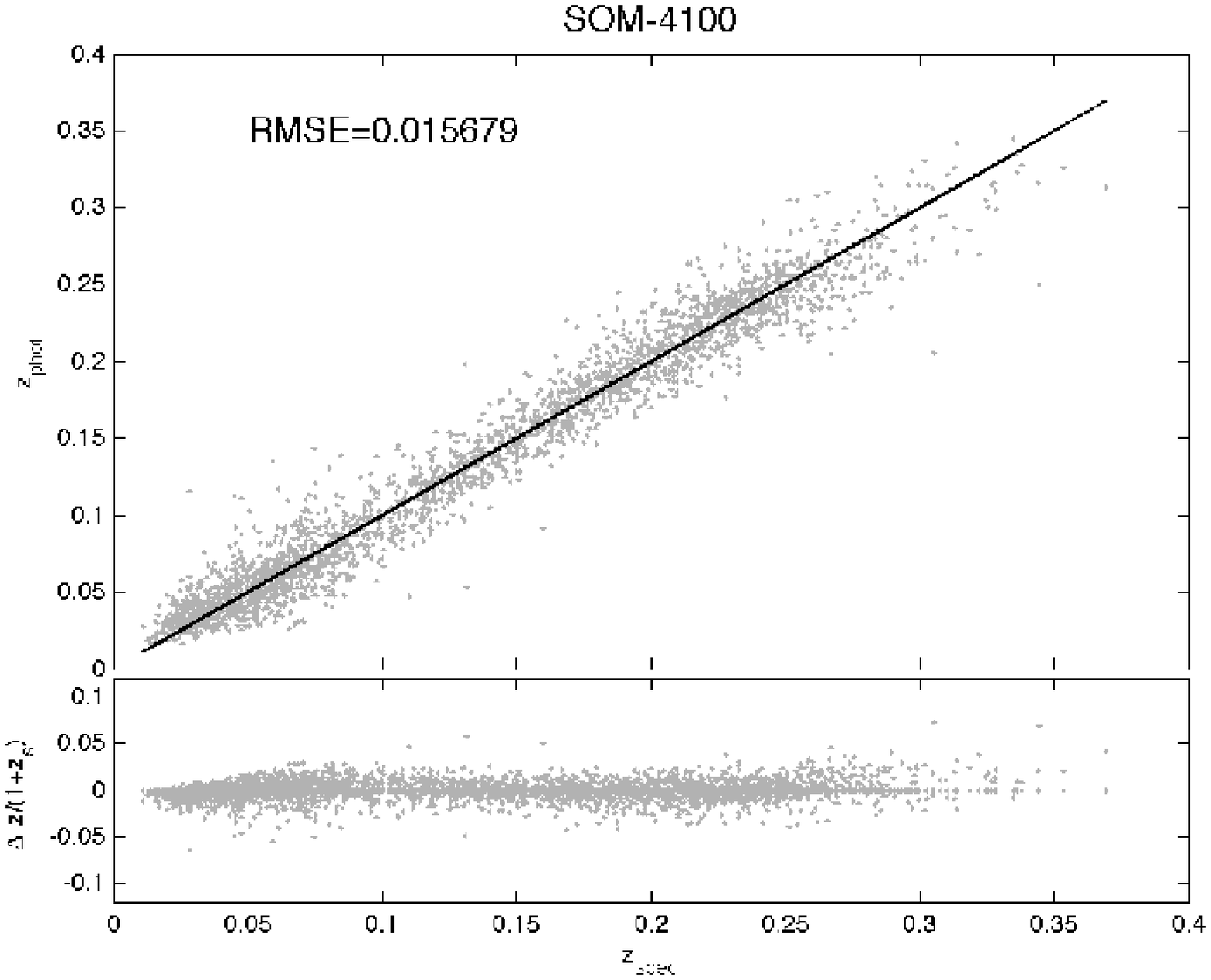} &
\includegraphics[scale=0.25]{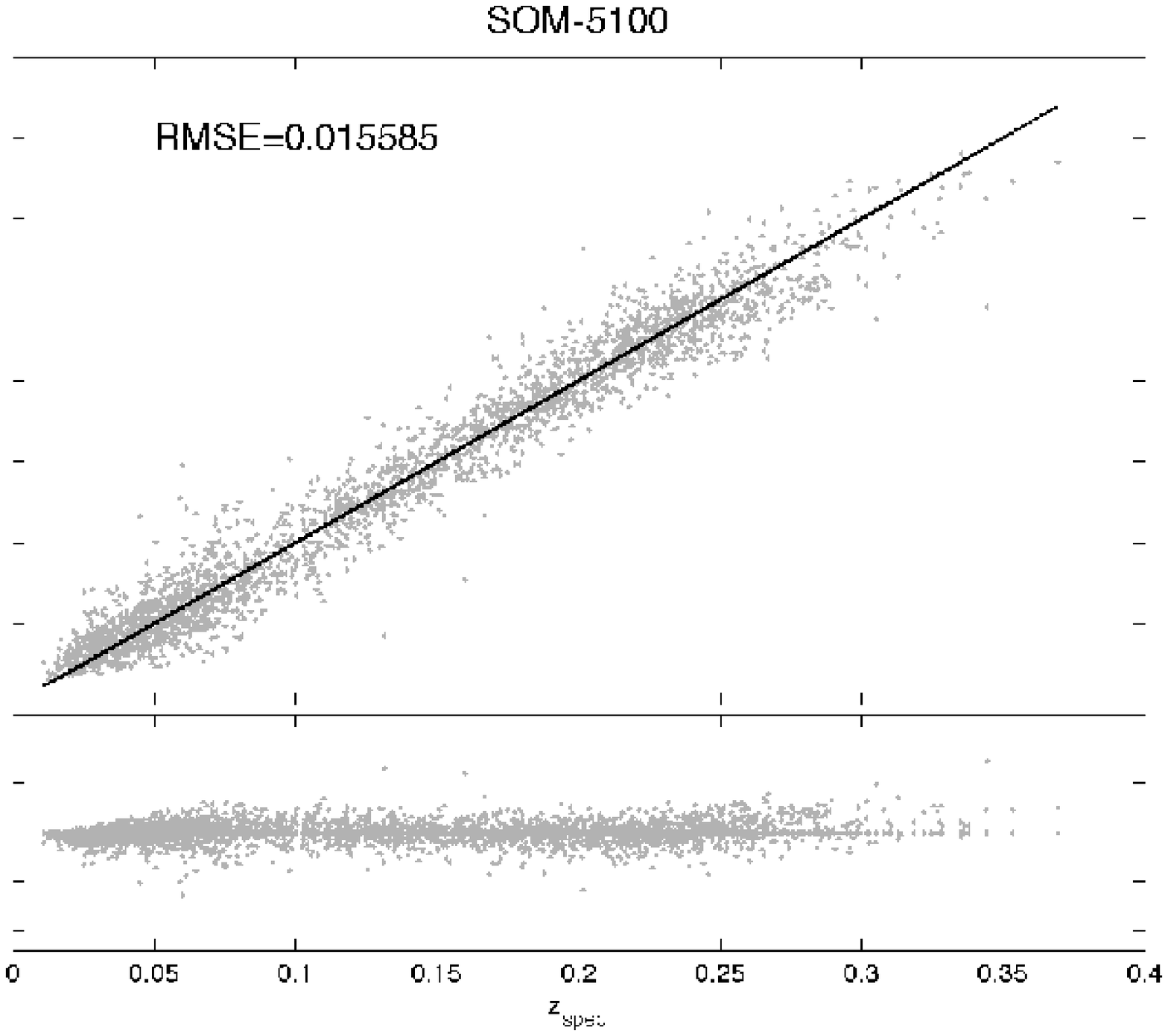} \\
\end{tabular}
\caption{Results from the three methods using SDSS u--g--r--i--z dereddened
magnitudes as inputs for the SDSS DR7 Luminous Red Galaxies classified
as ellipticals by the GalaxyZoo team. The bottom two plots show the SOM
results for the two local minima described in Section \ref{sec:methods} and shown in
Figure \ref{fig2}}\label{fig3}
\end{figure}

\section{Conclusion}\label{sec:conclusion}

SOMs offer a competitive choice in terms of low RMSE, algorithm comprehension
(also see \cite{GoeppertRosenstiel1993}) and percentage of outliers.
The final results are presented in Table \ref{tbl-1} and plots for
the LRG--ELL data set for the SOM, ANNz and GPR methods are shown in Figure \ref{fig3}

As mentioned previously, obtaining the global minimum is important
and, not surprisingly, can affect the results. Figure \ref{fig2}
shows the two local minima ($\xi$=4100 and 5100) listed for the LRG--ELL
(Luminous Red Galaxies classified as Ellipticals by GalaxyZoo)
data set in Table \ref{tbl-1}. Clearly there are a number of other $\xi$--values
and the RMSE will be greatly affected by the choice as seen on the y--axis of
Figure \ref{fig2} for a given $\xi$--value. Given these facts,
the roughness of the RMSE cost function in Figure \ref{fig2} shows that
traditional gradient based optimization strategies, e.g, deterministic
annealing, might yield sub--optimal solutions.  Other methods, such as, genetic
programming might find the ``global" minimum much faster, 
if a global minima exists with respect to the uncertainties of the RMSE.

During completion of this manuscript another paper using
SOMs for classification and photometric estimation was released
\citep{geach2011}. Our work differs in that we mostly focus on a wider variety
of low--redshift samples drawn from the SDSS, while \citep{geach2011}
focuses more on the higher redshift samples akin to those used in \cite{Hildebrandt2010}.

We have shown that SOMs are a powerful tool for estimating photometric
redshifts and that with additional work they are sure to be useful
in future surveys with limited numbers of follow--up spectroscopic redshifts.

\acknowledgments
M.J.W would like to thank the Astrophysics Department at Uppsala University
for their generous hospitality while part of this work was completed.

C.D.K. thanks Think Geohazards for providing the computational resources
needed for estimating photometric redshifts via Self--Organizing Mapping.

Thanks goes to Joe Bredekamp and the NASA Applied Information
Systems Research Program for support and encouragement.

Funding for the SDSS has been provided by
the Alfred P. Sloan Foundation, the Participating Institutions, the National
Aeronautics and Space Administration, the National Science Foundation,
the U.S. Department of Energy, the Japanese Monbukagakusho, and the Max
Planck Society. The SDSS Web site is http://www.sdss.org/.

The SDSS is managed by the Astrophysical Research Consortium for
the Participating Institutions. The Participating Institutions are The
University of Chicago, Fermilab, the Institute for Advanced Study, the
Japan Participation Group, The Johns Hopkins University, Los Alamos National
Laboratory, the Max--Planck--Institute for Astronomy, the
Max--Planck--Institute for Astrophysics, New Mexico State University,
University of Pittsburgh, Princeton University, the United States Naval
Observatory, and the University of Washington.

This research has made use of NASA's Astrophysics Data System Bibliographic
Services.

This research has also utilized the viewpoints \citep{GLW2010} software package.




\begin{thebibliography}{}

\bibitem[Abazajian et al.(2009)]{DR7}
Abazajian, K.N. et al. 2009, \apjs, 182, 543

\bibitem[Abdalla et al.(2011)]{Abdalla2011}
Abdalla, F.B., Banerji, M., Lahav, O. \& Rashkov, V. 2011, \mnras, 417, 1891

\bibitem[Arnouts et al.(2002)]{Arnouts2002}
Arnouts, S., Moscardini, L., Vanzella, E., et al. 2002, \mnras, 329, 355

\bibitem[Ball et al.(2004)]{Ball04}
Ball, N.M., Loveday, J., Fukugita, M., Nakamura, O., Okamura, S.,
Brinkmann, J., \& Brunner, R.J. 2004, \mnras, 348, 1038

\bibitem[Ben\'itez(2000)]{Benitez2000}
Ben\'itez, N. 2000, \apj, 536, 571

\bibitem[Bonfield et al.(2010)]{Bonfield2010}
Bonfield, D.G., Sun, Y., Davey, N., Jarvis, M.J., Abdalla, F.B.,
Banerji, M., \& Adams, R. G. 2010, \mnras, 405, 987

\bibitem[Boulade et al.(2003)]{Boulade2003}
Boulade, O., Charlot, X., Abbon, P., et al. 2003, ed. M. Iye, \& A. F. M.
Moorwood, SPIE Conf. Ser., 4841, 72

\bibitem[Carliles et al.(2008)]{Carliles08}
Carliles, S., et al. 2008, ASPC, 394, 521

\bibitem[Collister \& Lahav(2004)]{CL2004}
Collister, A. A. \& Lahav, O. 2004, \pasp, 116, 345

\bibitem[Csabai et al.(2003)]{Csabai03}
Csabai, I., et al. 2003, \aj, 125, 580

\bibitem[Eisenstein et al.(2001)]{Eisenstein01}
Eisenstein et al. 2001, \aj, 122, 2267

\bibitem[Firth et al.(2003)]{Firth03}
Firth, A.E., Lahav, O., \& Somerville, R.S. 2003, \mnras, 339, 1195

\bibitem[Foster et al.(2009)]{Foster09}
Foster, L., Waagen, A., Aijaz, N. et al. 2009, Journal of Machine Learning Research, 10, 857

\bibitem[Gazis, Levit, \& Way(2010)]{GLW2010}
Gazis, P.R., Levit, C. \& Way, M.J. 2010, \pasp, 122, 1518

\bibitem[Geach(2011)]{geach2011}
Geach, J.E. 2011, arXiv:1110.0005, \mnras in Press

\bibitem[G\"{o}ppert \& Rosenstiel(1993)]{GoeppertRosenstiel1993}
G\"{o}ppert, J. \& Rosenstiel, W. 1993, ``Self-organizing Maps vs.
Backpropagation: An Experimental Study'', Proc. of Workshop on Design
Methodologies for Microelectronis and Signal Processing, pp. 153--162,
Giwice, Poland.

\bibitem[Haykin(2009)]{Haykin2009}
Haykin, S.S. 2009, ``Neural networks and learning machines'', v.10,
Prentice Hall, ISBN 9780131471399

\bibitem[Hildebrandt et al.(2010)]{Hildebrandt2010}
Hildebrandt et al. 2010, \aap, 523, A31

\bibitem[Ilbert et al.(2006)]{Ilbert2006}
Ilbert, O., Arnouts, S., McCracken, H. J., et al. 2006, \aap, 457, 841

\bibitem[Ivezic et al.(2008)]{Ivezic08}
Ivezic, Z., Tyson, J.A., Allsman, R., Andrew, J., Angel, R., et al 2008,
arXiv:0805.2366v1

\bibitem[Kaiser(2004)]{Kaiser2004}
Kaiser, N. 2004, ``Pan-STARRS: a wide-field optical survey telescope array”,
SPIE, 5489, 11-12

\bibitem[Klose(2006)]{Klose2006}
Klose, C.D. 2006, Computational Geosciences; 10(3), 265-277

\bibitem[Klose et al.(2008)]{Klose-etal2008}
Klose, C.D., A.D., Netz, U., Scheel, A.K., Beuthan, J., Hielscher, A.H.
Biomed Opt 13(5):050503

\bibitem[Klose et al.(2010)]{Klose-etal2010}
Klose, C.D., A.D., Netz, U., Scheel, A.K., Beuthan, J., Hielscher, A.H. 2010,
Biomed Opt 15(6):066020

\bibitem[Kohonen(1982)]{Kohonen1982}
Kohonen, T. 1982 Biol. Cyb., 43(1): 59-69

\bibitem[Kohonen(2001)]{Kohonen2001}
Kohonen, T. 2001, Self--Organizing Maps, 3rd edition, Springer, Berlin.

\bibitem[Li \& Yee(2008)]{LiYee2008}
Li, I.H. \& Yee, H.K.C. 2008, \aj, 135, 809

\bibitem[Lintott et al.(2008)]{Lintott2008}
Lintott, C., Schawinski, K., Slosar, A. et al. 2011, \mnras, 389, 1179

\bibitem[Lintott et al.(2011)]{Lintott2011}
Lintott, C., Schawinski, K., Bamford, S. et al. 2011, \mnras, 410, 166

\bibitem[Mahdi(2011)]{Mahdi2011}
Mahdi, B. 2011, arXiv:1108.0514

\bibitem[Naim et al.(1997)]{Naim1997}
Naim, A., Ratnatunga, K.U. \& Griffiths, R.E. 1997, \apjs, 111, 357

\bibitem[Schneider et al.(2007)]{Schneider2007}
Schneider, D.P., Hall, P.B., Richards, G.T. et al. 2007, \aj, 134, 102

\bibitem[Sorba \& Sawicki(2011)]{Sorba2011}
Sorba, R. \& Sawicki, M. 2011, arXiv:1101.4635

\bibitem[Strauss et al.(2002)]{Strauss02}
Strauss, M.A., et al. 2002, \aj, 124, 1810

\bibitem[Suchkov et al.(2005)]{Suchkov05}
Suchkov, A.A., Hanisch, R.J., \& Margon, B. 2005, \aj, 130, 2439

\bibitem[Tagliaferri et al.(2003)]{Tagliaferri03}
Tagliaferri, R., Longo, G., Andreon, S., Capozziello, S., Donalek, C.,
\& Giordano, G. 2003, Lecture Notes in Computer Science, vol 2859, 226

\bibitem[Vanzella et al.(2004)]{Vanzella04}
Vanzella, E., et al. 2004, \aap, 423, 761

\bibitem[Wadadekar(2005)]{Wadadekar05}
Wadadekar, Y. 2005, \pasp, 117, 79

\bibitem[Way, Gazis \& Scargle(2011)]{WGS2011}
Way, M.J., Gazis, P.R. \& Scargle, J.D. 2011, \apj, 727, 48

\bibitem[Way \& Srivastava(2006)]{Way06}
Way, M.J. \& Srivastava, A.N. 2006, \apj, 647, 102

\bibitem[Way et al.(2009)]{Way09}
Way, M.J., Foster, L.V., Gazis, P.R. \& Srivastava, A.N. 2009, \apj, 706, 623

\bibitem[Way(2011)]{Way2011}
Way, M.J. 2011, \apjl, 734, 9

\bibitem[York et al.(2000)]{York2000}
York, D.G., et al. 2000, \aj, 120, 1579

\end{thebibliography}
\end{document}